\documentclass[aps,prl,nofootinbib,twocolumn]{revtex4}
\usepackage{amssymb}
\usepackage{amsmath,color}
\usepackage{epsfig}
\usepackage{graphicx,epsf,epsfig}
\usepackage{bbm}
\usepackage{subfigure}
\usepackage{multirow}
\usepackage{setspace}
\usepackage{verbatim}
\usepackage{epstopdf}
\newcommand{\be}{\begin{eqnarray}}
\newcommand{\ee}{\end{eqnarray}}

\begin{document}

\title{Gravitational Waves from Oscillons with Cuspy Potentials}

\author{Jing Liu$^{1,2}$}
\email{liujing@itp.ac.cn}

\author{Zong-Kuan Guo$^{1,2}$}
\email{guozk@itp.ac.cn}

\author{Rong-Gen Cai$^{1,2}$}
\email{cairg@itp.ac.cn}

\author{Gary Shiu$^{3,4}$}
\email{shiu@physics.wisc.edu}

\affiliation{$^1$CAS Key Laboratory of Theoretical Physics, Institute of Theoretical Physics,
 Chinese Academy of Sciences, P.O. Box 2735, Beijing 100190, China}
\affiliation{$^2$School of Physical Sciences, University of Chinese Academy of Sciences,
 No.19A Yuquan Road, Beijing 100049, China}
\affiliation{$^3$Department of Physics, University of Wisconsin-Madison, Madison, WI 53706, USA}
\affiliation{$^4$International College, University of Chinese Academy of Sciences, Beijing, China}

\begin{abstract}
We study the production of gravitational waves during oscillations of the inflaton
around the minimum of a cuspy potential after inflation.
We find that a cusp in the potential can trigger copious oscillon formation,
which sources a characteristic energy spectrum of gravitational waves with double peaks.
The discovery of such a double-peak spectrum could test the underlying inflationary physics.
\end{abstract}

\maketitle

\emph{Introduction}. Gravitational waves (GWs) play an important role in the context of inflationary cosmology.
A stochastic background of GWs,
produced during inflation and subsequent preheating/reheating after inflation,
carries useful information about the inflationary dynamics and
inflaton decay (see~\cite{Cai:2017cbj} for a recent review).
Detecting such a stochastic background of GWs, whether directly or indirectly,
can provide us with a unique opportunity to test theories of inflation.

During inflation, quantum fluctuations of the tensor modes of the spacetime metric were stretched
by the accelerated expansion of the Universe, and were then nearly frozen on super-Hubble scales.
Since these GWs can result in B-mode polarization of the cosmic microwave background (CMB) anisotropies,
their spectrum is in principle measurable by CMB polarization experiments.
Current CMB data alone already put an upper bound on the tensor-to-scalar ratio $r < 0.09$ at 95$\%$ confidence level \cite{Array:2015xqh}, and when combined with the constraints on the scalar spectral index, have been effective in discriminating inflationary models.
For example, the cubic and quartic potentials are strongly disfavored,
and the quadratic potential is moderately disfavored by the Planck 2015 data~\cite{Ade:2015lrj}, while axion monodromy inflation with a linear potential~\cite{McAllister:2008hb}
or fractional powers~\cite{Silverstein:2008sg} are compatible with the current Planck results.
Further advances in axion monodromy inflation have suggested potentials with even more possible powers \cite{Marchesano:2014mla,McAllister:2014mpa}. Moreover, it has recently been shown  that stringy effects
can lower the power of a quadratic axion monodromy potential to less than linear  \cite{Landete:2017amp}.
Thus, axion monodromy inflation represents an interesting class of large field inflationary models that are compatible with data.

Besides vacuum fluctuations during inflation, another source of GWs is parametric resonance during preheating \cite{Kofman:1994rk}.
During preheating after inflation,
the Fourier modes of a scalar matter field $\chi$ coupled to the inflaton
grow exponentially by parametric resonance, driven by the oscillating inflaton.
The resonant modes are quickly pumped up to a large amplitude.
Such highly pumped modes correspond to large, time dependent density inhomogeneities in position space,
ensuring that the matter distribution has a non-trivial quadrupole moment,
which can source a significant spectrum of GWs~\cite{Khlebnikov:1997di}.
The present peak frequency of such GWs is proportional to
the energy scale of inflation~\cite{Easther:2006gt},
while the present amplitude of GWs
is independent of the energy scale of inflation~\cite{Easther:2006vd}.
In hybrid inflation
the stochastic background produced during preheating
is expected to be directly detected by future GW detectors~\cite{GarciaBellido:2007dg}.

In this Letter, we shall investigate the production of GWs during oscillations of the inflaton
after inflation with a cuspy potential
\be
V(\phi)=\lambda M_\mathrm{pl}^{4-p}|\phi|^p,
\label{potentials}
\ee
with $p=1, 2/3, 2/5$, and $M_\mathrm{pl}\equiv (8\pi G)^{-1/2}$ is the reduced Planck mass.
In string theory, axion monodromy can be introduced by space-filling wrapped branes
leading to a linear potential~\cite{McAllister:2008hb}.
Inflationary potentials with powers of $2/3$ and $2/5$ arise in compactifications on manifolds
with metric flux such as Nil manifolds which contain tori twisted over circles~\cite{Silverstein:2008sg}.
More generally, monodromy generated by fluxes can lead to potentials with more varieties of power~\cite{Marchesano:2014mla, McAllister:2014mpa}.
{
Here we hasten to add that the powers of these potentials are expected only at large field values, due to the
coupling of the inflaton to high scale physics.
At the end of inflation, i.e., for small $\phi$, these potentials for axion monodromy become quadratic. Nonetheless, cuspy potentials can arise in other inflationary contexts, e.g., through non-standard kinetic terms or as a result of integrating out the dynamics of other fields that couple to the inflaton. Thus, we use these specific potentials as benchmarks to illustrate the
novel GW signatures that can arise when the potential has a cuspy behavior at the end of inflation.
Assuming the potential in eq.~(\ref{potentials}) applies to both the inflationary era and at the end of inflation,
the value of $\lambda$
in this simple class of models
can be fixed by the estimated amplitude of scalar perturbations
from the CMB data.
For powers of $p=1, 2/3, 2/5$, $\lambda \approx 3, 4, 5\times10^{-10}$,
the predicted scalar spectral index $n_s\approx 0.970, 0.973, 0.976$,
and the predicted tensor-to-scalar ratio $r\approx 0.08, 0.05, 0.03$,
respectively, assuming the number of e-folds $N=50$.
These predictions are in agreement with the recent CMB data.
In the reheating scenario, the inflaton $\phi$ oscillates near the minimum of its potential after inflation
and decays into elementary particles.
However, due to the cusp of the potential, the oscillating behavior of the inflaton is very different from that of smooth potentials like $\phi^2$ and $\phi^4$.
It has recently been shown that an efficient parametric resonance can occur
during preheating for an inflaton potential of the form of eq.~(\ref{potentials}) with $0<p \leq 2$,
if the inflaton is coupled to a scalar matter field $\chi$ via an interaction term $\phi^2\chi^2$~\cite{Moghaddam:2015ava}.
However, the production of GWs has not been studied to our knowledge.
In this Letter, we are interested in the production of GWs
during oscillations of the inflaton after inflation with cuspy potentials.
We find that the non-smooth oscillations can trigger
amplification of fluctuations of the inflaton itself at the moment when $\phi(t)=0$,
so that oscillons copiously form after inflation.
As in the models with a symmetric smooth potential~\cite{Zhou:2013tsa}
and an asymmetric smooth potential~\cite{Antusch:2016con},
the oscillon formation in the models with cuspy potentials
sources a stochastic background of GWs,
on which the characteristic size of the oscillons is imprinted.
Interestingly, these cuspy potentials result in a characteristic energy spectrum of GWs with double peaks,
which can be distinguished from smooth potentials by probing the shape of the energy spectrum of GWs.

\emph{Model}.
GWs are described by the transverse-traceless gauge-invariant tensor perturbation,
$h_{ij}$, in a Friedman-Robertson-Walker (FRW) metric,
\be
ds^2=-dt^2+a^2(t)(\delta_{ij}+h_{ij})dx^idx^j.
\ee
The perturbed Einstein equation reads
\be
\ddot{h}_{ij}+3H\dot{h}_{ij}-\frac{1}{a^2}\nabla^2 h_{ij}=\frac{2}{M_\mathrm{pl}^2a^2}\Pi_{ij}^\mathrm{TT},
\label{eq:gwh}
\ee
where $\Pi_{ij}^\mathrm{TT}$ is the transverse-traceless projection of the anisotropic stress tensor $T_{ij}$.
In our model we assume that the inflaton is weakly coupled to other fields during preheating.
GWs are sourced by the inflaton $\phi$,
i.e., $\Pi_{ij}^\mathrm{TT}=(\partial_i \phi\partial_j \phi)^\mathrm{TT}$.
The energy density of GWs is
\be
\rho_\mathrm{GW}=\frac{M_\mathrm{pl}^2}{4}\langle\dot{h}_{ij}\dot{h}^{ij}\rangle,
\ee
where $\langle...\rangle$ denotes a spatial average over the volume.
It is commonly parameterized by the dimensionless density parameter per logarithmic frequency interval,
$\Omega_\mathrm{GW} = d\rho_\mathrm{GW}/d\ln k/\rho_c$,
where $\rho_c$ is the critical density of the Universe.
The energy density spectrum of the inflaton is defined as
\be
k^3 \rho_k = \frac12 k^3 \left(|\partial_\tau \varphi_k|^2 + \omega_k^2|\varphi_k|^2\right),
\ee
where $\omega_k=\sqrt{k^2 + a^2\langle V^{\prime\prime}\rangle - \partial_\tau^2 a/a}$,
$\varphi_k$ are the Fourier modes of $a\phi$, and $\tau$ is the conformal time.

To understand oscillon formation during oscillations of the inflaton after inflation with a cuspy potential,
we now investigate the evolution behavior of inflaton fluctuations.
To first order, the equation of motion for the Fourier modes of fluctuations of the inflaton $\phi$ is
\be
\ddot{\delta\phi_k}+3H\dot{\delta\phi_k}+\left(\frac{k^2}{a^2}+V''(\phi)\right)\delta\phi_k=0.
\ee
To solve this equation, we neglect the expansion of the Universe for the moment,
thus the friction term drops out of the equation of motion.
Moreover, since we are interested in large-scale modes,
the $k^2$ term can be dropped.
The equation becomes
\be
\ddot{\delta\phi_k}+V''(\phi)\delta\phi_k=0.
\ee
For illustrative purposes, in what follows let us consider the linear potential.
Since the inflaton potential $V(\phi)=\lambda M_\mathrm{pl}^3|\phi|$ has a cusp at $\phi=0$,
its derivative with respect to $\phi$ is a step function
and its second order derivative is a delta function $V''(\phi)=2\lambda M_\mathrm{pl}^3 \delta(\phi)$.
We focus on the evolution behavior of the $\delta \phi_k$ modes near the point $\phi(t)=0$.
It is convenient to define $t$ such that $\phi(t=0)=0$.
The solution to the equation of motion for the inflaton is
$\phi(t)=|\dot{\phi}_m|t+\lambda M_\mathrm{pl}^3 t^2/2$ when $t<0$
and $\phi(t)=|\dot{\phi}_m|t-\lambda M_\mathrm{pl}^3 t^2/2$ when $t>0$,
where $\dot{\phi}_m$ is the maximum value of $\dot{\phi}$ at $t=0$.
Since $\phi(t) \approx |\dot{\phi}_m|t$ is a good approximation in a small vicinity of $\phi=0$,
we find
$\delta \dot{\phi}_k(t=0^+)-\delta \dot{\phi}_k(t=0^-)=-2\lambda M_\mathrm{pl}^3 \delta \phi_k(t=0)/|\dot{\phi}_m|$,
which implies that $\delta \dot{\phi}_k$ jumps suddenly when $\phi$ crosses the cusp of the potential.
Such periodic jumps of $\delta\dot{\phi}_k$ result in periodic, rapid increases of the energy density $\rho_k$.
We show the time evolution of the energy density for a linear potential (orange)
in an expanding Universe in Fig.~\ref{fig:rhok}.
In the cases of $p=2/3$ and $p=2/5$, $|V'(\phi)|$ becomes infinite when $|\phi|$ tends to zero.
To avoid this singularity, the potential with a cut-off $|\phi|>{\cal O}(10^{-3})$ is adopted in our numerical calculations.
We also show the time evolution of the energy density for the $\phi^{2/3}$ potential (blue)
and $\phi^{2/5}$ potential (green) in Fig.~\ref{fig:rhok}.
Similar to the linear potential, the energy density suddenly increases near the points at which $\phi(t)=0$.
However, the sudden increase is followed by a sudden decrease near $\phi(t)=0$
after several oscillations of the field $\phi$.
The energy density always increases after each oscillation of the field $\phi$.
We have checked that the increase in energy density is independent of the choice of the cut-off.
As a result, oscillon formation occurs when the inflaton oscillates
near the minimum of its potential,
which sources a stochastic background of GWs.

\begin{figure}[h]
\includegraphics[width=3.2in]{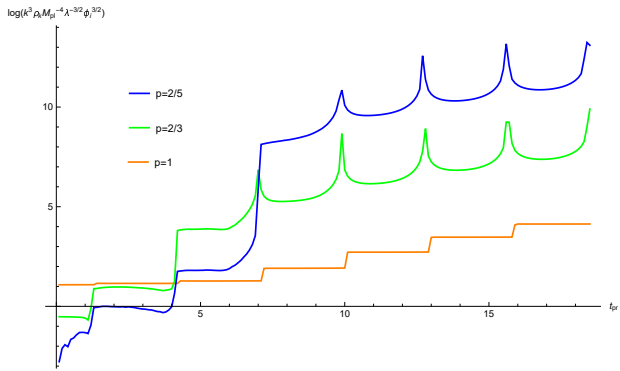}
\caption{Evolutions of $\rho_k$ for cuspy potentials with $p=1$ and $k=1.0\times 10^{-3}M_\mathrm{pl}$ (orange),
$p=2/3$ and $k=1.0\times 10^{-5}M_\mathrm{pl}$ (blue), $p=2/5$ and $k=1.3\times 10^{-4}M_\mathrm{pl}$ (green), respectively.}
\label{fig:rhok}
\end{figure}

\begin{figure*}[!]
\includegraphics[width=3.2in]{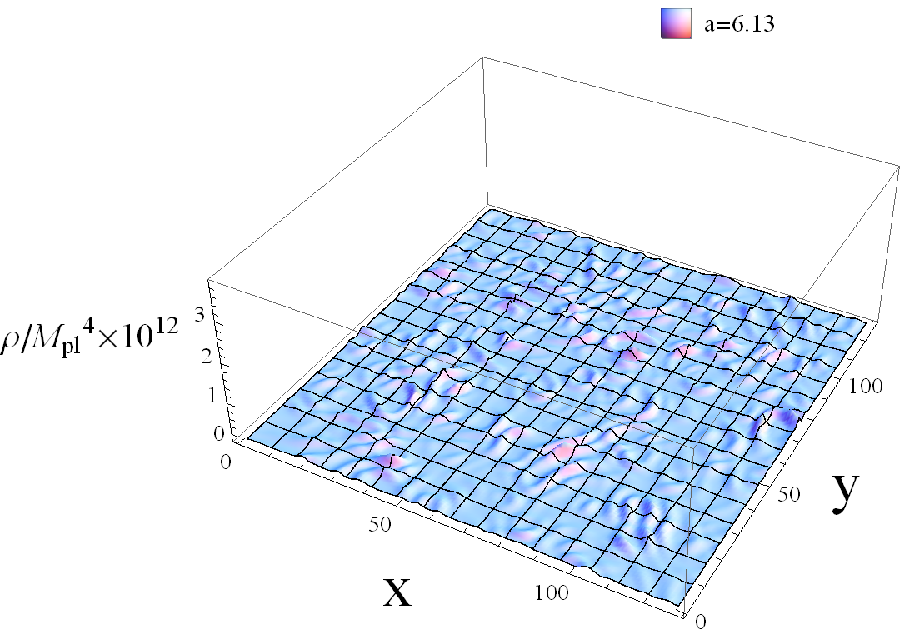}
\includegraphics[width=3.2in]{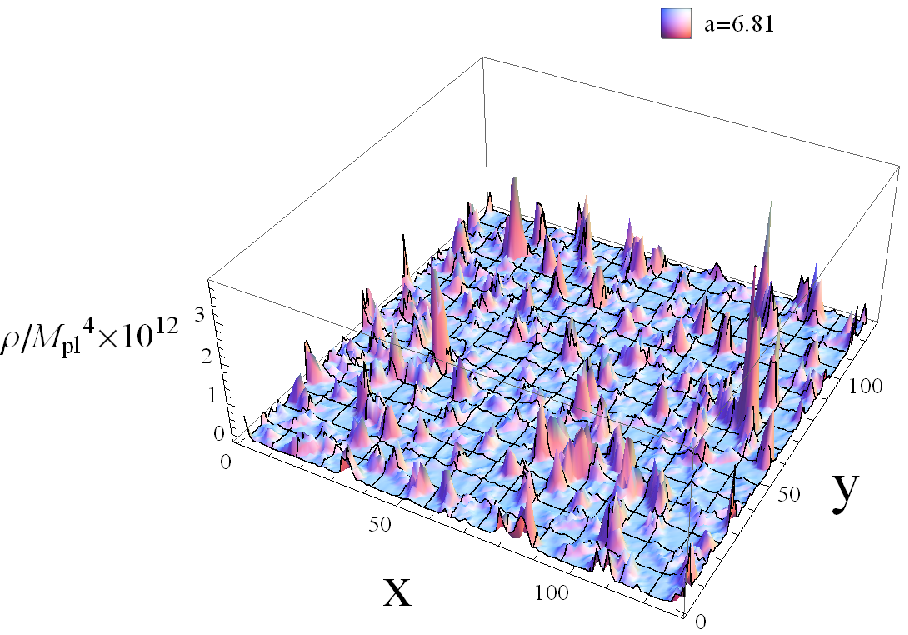}
\includegraphics[width=3.2in]{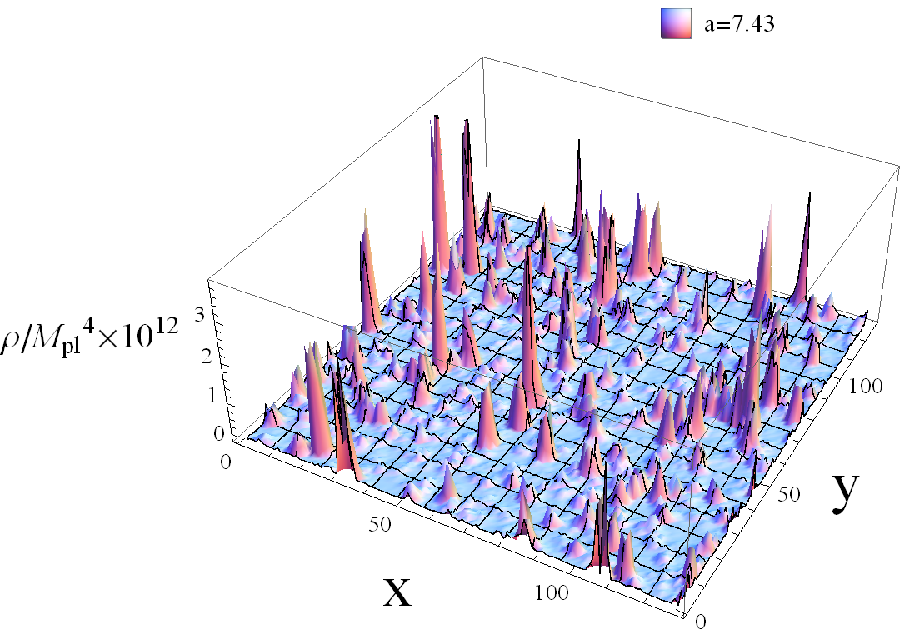}
\includegraphics[width=3.2in]{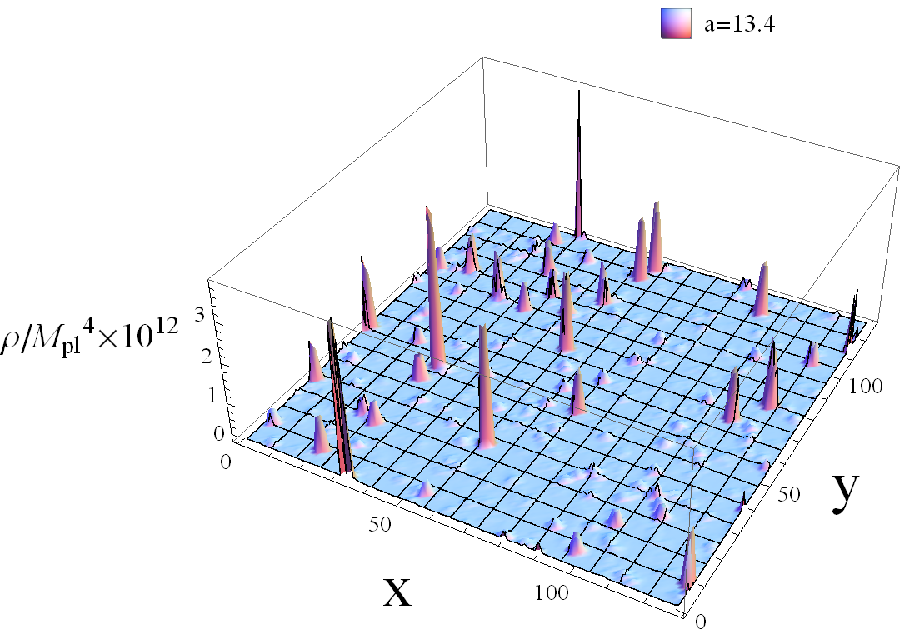}
\caption{Energy density $\rho$ on a two-dimensional slice through the simulation,
when $a(t)=6.13$ (top-left), $6.81$ (top-right), $7.43$ (bottom-left) and $13.4$ (bottom-right),
in the linear potential model.}
\label{fig:oscillon}
\end{figure*}

\emph{Simulation Results}.
Using a modified version of LATTICEEASY~\cite{Felder:2000hq},
we simulate the production of GWs during preheating in the models~\eqref{potentials}
with cuspy potentials.
LATTICEEASY has been developed for more generally calculating the evolution of interacting scalar fields in an expanding Universe.
The spectral method can directly solve
the GW equation~\eqref{eq:gwh}
in Fourier space~\cite{Easther:2007vj}.
Actually, one can first evolve the tensor perturbations in configuration space
and then apply the transverse-traceless projector to the real physical $h_{ij}$ in Fourier space~\cite{GarciaBellido:2007af}.
Another method is based on the Green's functions in Fourier space to calculate numerically the energy spectrum of GWs
generated well inside the horizon~\cite{Dufaux:2007pt}.
In our lattice simulations we adopt the configuration-space method for solving the following evolution equation of the tensor perturbations
\be
\ddot{u}_{ij}+3H\dot{u}_{ij}-\frac{1}{a^2}\nabla^2 u_{ij}=\frac{2}{M_\mathrm{pl}^2a^2}T_{ij}.
\label{eq:gwu}
\ee
Therefore, the transverse-traceless tensor perturbations can be written as
$h_{ij}(t,\mathbf{k})=\Lambda_{ij,lm}(\hat{\mathbf{k}})u_{lm}(t,\mathbf{k})$,
where $\Lambda_{ij,lm}(\hat{\mathbf{k}})$ is the transverse-traceless projection operator
and $u_{lm}(t,\mathbf{k})$ is the Fourier transform of the solution to the equation~\eqref{eq:gwu}.
The energy density of GWs can be expressed in terms of $u_{ij}$ as
\be
\rho_\mathrm{GW}=\frac{M_\mathrm{pl}^2}{4L^3}\int d^3\mathbf{k} \Lambda_{ij,lm}(\hat{\mathbf{k}})\dot{u}_{ij}(t,\mathbf{k})\dot{u}^\ast_{lm}(t,\mathbf{k}).
\ee

We perform three-dimensional lattice simulations with $256^3$ points in a box with periodic boundary conditions
assuming $\lambda=1.26 \times 10^{-12}$ in the linear potential model.
We set the initial values of the inflaton, its derivative and the scale factor as
$\phi_i=0.75 M_\mathrm{pl}$, $\dot{\phi}_i=6.8\times10^{-4}M_\mathrm{pl}^2$ and $a_i=1$.
The inflaton fluctuations and its derivative are initialized by quantum vacuum fluctuations,
while the tensor fluctuation and its derivative are initialized as zero.
We stop the simulation when the energy spectrum of GWs does not grow significantly.
We assume that reheating ends at the end of the simulation.
After that, the Universe enters into the radiation-dominated era.
The energy spectrum and its frequency at the end of simulations are converted to the present values.
Fig.~\ref{fig:oscillon} shows the time evolution of the energy density as a function
of position on a two-dimensional slice through the simulation
from $a(t)=6.13$ (top-left), to $6.81$ (top-right), to $7.43$ (bottom-left) and $13.4$ (bottom-right)
in the linear potential model.
We can see that at the beginning, oscillons copiously form and then decay during oscillations of the inflaton.
In this model, the rapid growth of oscillons results in the production of GWs
with $\Omega_\mathrm{GW}h^2 \sim 2 \times 10^{-9}$ today.
Our lattice simulation results show that there appear two peaks in the energy spectrum of GWs,
a feature very distinct from that of other models.
Therefore, our model can be distinguished from the production of GWs during preheating by future GW detectors.
While this paper was in preparation, a phenomenological study of GWs produced from preheating with a time dependent resonance parameter $q(t)$ was recently undertaken \cite{Figueroa:2017vfa}. For some choices of $q(t)$, one also finds a GW spectrum with multiple peaks due to non-linear effects.
The double peak of GWs  in \cite{Figueroa:2017vfa} arises due to the parametric resonance in the preheating phase,
while in our work, it is not the case, the double peak is due to  the copious oscillon formation triggered by the cusp in the potential.

The evolution of the spectrum goes through two different stages, the linear growth stage and nonlinear growth stage.
In the first stage, as shown in Fig.~\ref{fig:rho_k_t}, the
small-$k$ modes of the field $\phi$ exponentially grow due to the cusp of the potential until the turning point $a(t)=7.30$.
The linear growth is more efficient than those driven by
the symmetric potential~\cite{Zhou:2013tsa} and asymmetric potential~\cite{Antusch:2016con}.
This leads to the left peak in the energy spectrum of GWs,
which is characteristic of the cuspy potential.
In the second stage, from Fig.~\ref{fig:rho_k_t} we see that the small-$k$ modes begin to drop
and the large-$k$ modes continue to grow.
It implies that the energy flows from the small-$k$ modes to the large-$k$ modes,
as discussed in detail in~\cite{GarciaBellido:2007af}.
This leads to the right peak in the energy spectrum of GWs.

\begin{figure}[h!]
\includegraphics[width=3.2in]{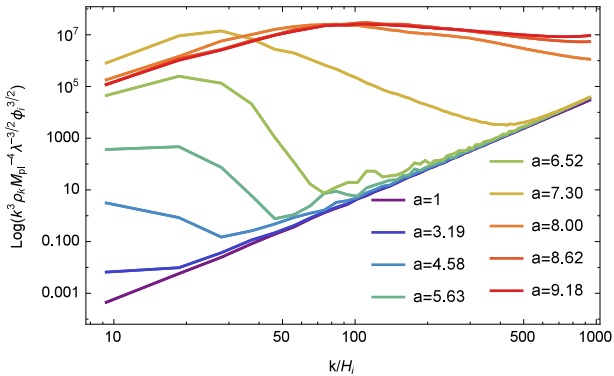}
\caption{Evolution of the energy density spectrum of the field $\phi$ for the linear potential.
The yellow line corresponds to a turning point $a(t)=7.30$.}
\label{fig:rho_k_t}
\end{figure}

\begin{figure}[h!]
\includegraphics[width=3.2in]{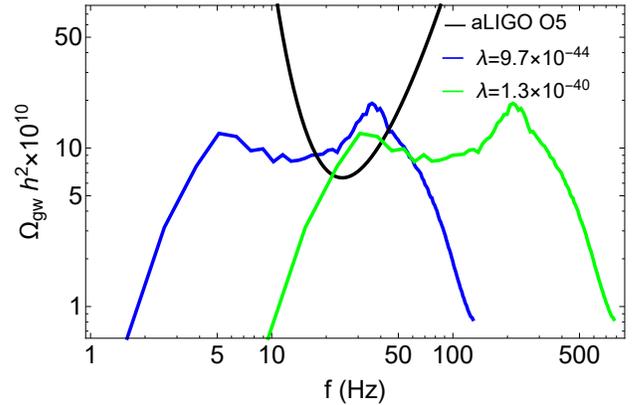}
\caption{Energy spectra of GWs today, predicted by the linear potential
with $\lambda=9.7\times 10^{-44}$ (blue) and $\lambda=1.3\times10^{-40}$ (green).
The black curve is the expected sensitivity curve of the fifth observing run (O5) of
the aLIGO-Virgo detector network.}
\label{fig:gwLIGO}
\end{figure}

Moreover, from the lattice simulations of preheating in the models~\eqref{potentials} with $p=2/3$ and $p=2/5$,
we find that the energy spectra of GWs peak at around
$\Omega_\mathrm{GW}h^2 \sim 1.2 \times 10^{-9}$ and $\Omega_\mathrm{GW}h^2 \sim 4 \times 10^{-10}$, respectively,
which are lower than the one in the linear potential model.

In our analysis, we have neglected the interactions between the inflaton $\phi$ and other matter fields.
If the inflaton is coupled to a matter field $\chi$,
broad parametric resonance leads effectively to a fast growth of the fluctuations of $\chi$~\cite{Moghaddam:2015ava}.
However, our numerical simulations confirm that the growth of the inflaton fluctuations themselves triggered by the cusp in its potential
is more effective than that of the field $\chi$ by parametric resonance.
Therefore, GWs are sourced mainly by the inflaton fluctuations,
even if a parametric resonance for the field $\chi$ occurs in this model.

\emph{Observational Implications}.
As found in~\cite{Easther:2006vd, GarciaBellido:2007dg},
the peak amplitude of the energy spectrum of GWs
needs not depend on the energy scale of inflation,
while the peak frequency scales inversely with the energy scale of inflation.
In the single-field slow-roll inflationary scenario, if $\lambda \approx 3\times10^{-10}$
is fixed by the amplitude of the primordial 
curvature perturbation
$A_s=2.2\times10^{-9}$~\cite{Ade:2015lrj},
the peak frequency of GWs today is fixed to  be $f \sim 10^{9}$Hz,
many orders of magnitude beyond the frequencies that can be reached by current
GW detection experiments.

If the model parameter $\lambda$ is not fixed by the amplitude of the primordial 
curvature perturbation,
the sensitivity of advanced LIGO (aLIGO) is expected to be significantly improved,
which allows us to possibly observe GWs produced during oscillations of inflaton after inflation.
For example, in the hybrid inflationary scenario~\cite{Linde:1993cn},
since $\phi$ is not necessarily the inflaton itself,
$\lambda$ becomes essentially a free parameter.
In this case we have plotted in Fig.~\ref{fig:gwLIGO} the present-day energy spectra of GWs
produced during oscillon formation in the linear potential model~\eqref{potentials}
with $\lambda=9.7\times 10^{-44}$ (blue) and $\lambda=1.3\times10^{-40}$ (green),
the peaks of which lie above the expected sensitivity curve of the fifth observing run (O5) of
the aLIGO-Virgo detector network~\cite{TheLIGOScientific:2016wyq}.
Fig.~\ref{fig:gwLIGO} shows that there are two peaks in the energy spectrum of GWs,
which differ from other spectra of GWs produced during preheating.
A detection of the second peak may require corroboration from other GW detectors such as the Big Bang Observatory.

To summarize, we have studied the production of GWs during oscillon formation
after inflation with cuspy potentials.
At the end of inflation, oscillon formation can be triggered
by the particular oscillations of the inflaton around the minimum of its potential,
which sources a characteristic double-peak spectrum of GWs.
The discovery of such a background would open a new observational window into inflationary physics.

\begin{acknowledgments}
This work is supported in part by the National Natural Science Foundation of China Grants
No.11690021, No.11690022, No.11575272, No.11335012, No.11375247 and No.11435006,
in part by the Strategic Priority Research Program of the Chinese Academy of Sciences Grant No. XDB23030100
and by Key Research Program of Frontier Sciences, CAS.
GS is supported in part by the DOE grant DE-SC0017647 and the Kellett Award of the University of Wisconsin.
\end{acknowledgments}

\end{document}